\documentclass[letterpaper]{article} 
\usepackage{aaai2026}  
\usepackage{times}  
\usepackage{helvet}  
\usepackage{courier}  
\usepackage[hyphens]{url}  
\usepackage{graphicx} 
\usepackage{natbib}  
\usepackage{caption} 
\usepackage{algorithm}
\usepackage{algorithmic}

\usepackage{newfloat}
\usepackage{listings}
\DeclareCaptionStyle{ruled}{labelfont=normalfont,labelsep=colon,strut=off} 
\floatstyle{ruled}
\newfloat{listing}{tb}{lst}{}
\floatname{listing}{Listing}
\usepackage{xcolor}
\usepackage[most]{tcolorbox}
\usepackage{lipsum}
\usepackage{amsfonts}
\usepackage{booktabs}  
\usepackage{multirow}  
\usepackage{subcaption}

\definecolor{lightblue}{rgb}{0.804, 0.902, 0.922}
\definecolor{lightpink}{rgb}{0.98,0.84,0.86}
\definecolor{customgreen}{HTML}{A8D08D}
\definecolor{customyellow}{HTML}{FFD966}

\title{Quantifying the Potential to Escape Filter Bubbles: \\ A Behavior-Aware Measure via Contrastive Simulation}
\author{
    Difu Feng\textsuperscript{\rm 1}\textsuperscript{\rm 2},
    Qianqian Xu\textsuperscript{\rm 1}\thanks{Corresponding author.},
    Zitai Wang\textsuperscript{\rm 1},
    Cong Hua\textsuperscript{\rm 1}\textsuperscript{\rm 2},
    Zhiyong Yang\textsuperscript{\rm 2},
    Qingming Huang\textsuperscript{\rm 2}\textsuperscript{\rm 3}\textsuperscript{\rm 1}\footnotemark[1]
}
\affiliations{
    \textsuperscript{\rm 1} Institute of Computing Technology, Chinese Academy of Sciences \\
    \textsuperscript{\rm 2} School of Computer Science and Technology, University of Chinese Academy of Sciences\\
    \textsuperscript{\rm 3} Key Laboratory of Big Data Mining and Knowledge Management, University of Chinese Academy of Sciences\\
    \{fengdifu24, yangzhiyong21, qmhuang\}@ucas.ac.cn, \{xuqiangian, wangzitai, huacong23z\}@ict.ac.cn 

}

\usepackage{bibentry}

\begin{document}

\maketitle

\begin{abstract}
Nowadays, recommendation systems have become crucial to online platforms, shaping user exposure by accurate preference modeling. However, such an exposure strategy can also reinforce users’ existing preferences, leading to a notorious phenomenon named filter bubbles. Given its negative effects, such as group polarization, increasing attention has been paid to exploring reasonable measures to filter bubbles. However, most existing evaluation metrics simply measure the diversity of user exposure, failing to distinguish between algorithmic preference modeling and actual information confinement. In view of this, we introduce Bubble Escape Potential ($\mathsf{BEP}$), a behavior-aware measure that quantifies how easily users can escape from filter bubbles. Specifically, $\mathsf{BEP}$ leverages a contrastive simulation framework that assigns different behavioral tendencies (e.g., positive vs. negative) to synthetic users and compares the induced exposure patterns. This design enables decoupling the effect of filter bubbles and preference modeling, allowing for more precise diagnosis of bubble severity.
We conduct extensive experiments across multiple recommendation models to examine the relationship between predictive accuracy and bubble escape potential across different groups. To the best of our knowledge, our empirical results are the first to quantitatively validate the dilemma between preference modeling and filter bubbles. What's more, we observe a counter-intuitive phenomenon that mild random recommendations are ineffective in alleviating filter bubbles, which can offer a principled foundation for further work in this direction.

\begin{links}
    \link{Code}{https://github.com/fengdifu24/bepmetric}
\end{links}

\end{abstract}

\begin{figure*}[!ht]
  \centering
  \includegraphics[width=0.95\textwidth]{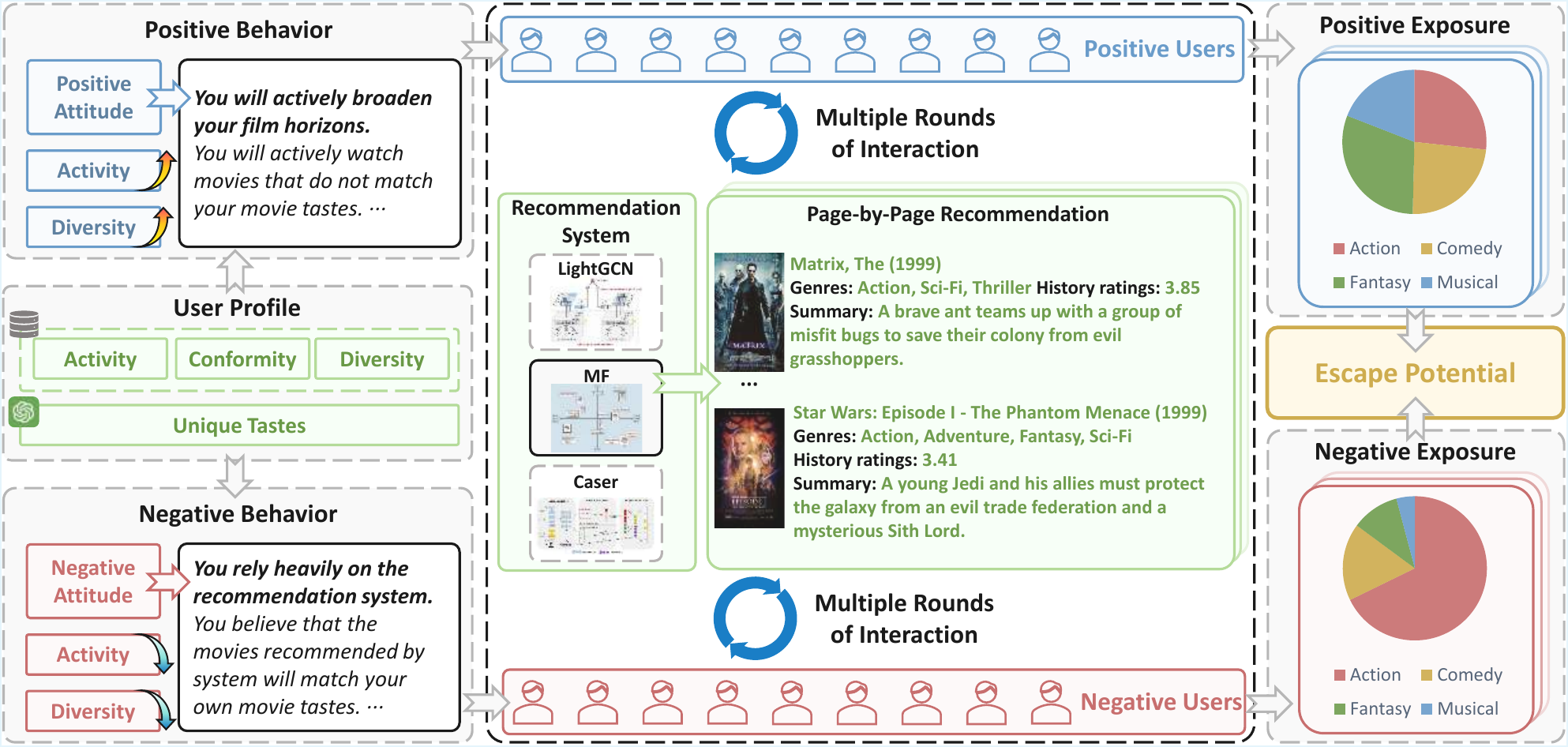}
  \caption{\textbf{Overview of the Bubble Escape Potential ($\mathsf{BEP}$) evaluation framework.}
We simulate two groups of users—\textbf{positive} and \textbf{negative}—who share user profiles (conformity and tastes). By modifying the user's activity, diversity and different attitudes in prompts, we generate positive users and negative users. Both groups interact with a selected recommendation system across multiple rounds, receiving page-by-page recommendations and making choices. By comparing their exposure, we quantify the system's \textbf{Bubble Escape Potential ($\mathsf{BEP}$)}.}
  \label{fig:procedure}
\end{figure*}

\section{Introduction}

Recommendation systems have become an integral part of online platforms, shaping how users access information in domains such as e-commerce, social media, and news. By modeling user preferences and tailoring content accordingly, these systems help users cope with overwhelming choices. However, this personalization comes at a cost: it often reinforces users’ existing interests and behaviors, leading to a well-known phenomenon called the \textit{filter bubble} \cite{pariser2011filter}. Within a filter bubble, users are repeatedly exposed to similar content, which may gradually restrict their worldview, amplify bias, and contribute to societal polarization \cite{bakshy2015exposure, ledwich2019algorithmic, han2025lightfair}.

Given the potential harm of filter bubbles, there has been growing interest in methods to detect and mitigate them. A common approach is to evaluate the diversity of content a user is exposed to, using metrics such as category count, coverage, or entropy \cite{gao2023cirs, piao2023human, ming2024modeling}. However, most existing methods focus solely on item-side properties and overlook a crucial aspect: user behavior. Recommendation systems are fundamentally interactive—the outcome is shaped not only by the algorithm but also by the user's own actions and preferences. Therefore, solely relying on content diversity fails to distinguish between algorithmic bias and natural user preference.

To incorporate user behavior into the measurement of filter bubble severity, we propose a new insight: \textbf{a filter bubble is more severe when a user actively tries to escape it but still fails to see diverse content.} In this light, we propose a novel, behavior-aware metric named \textbf{Bubble Escape Potential ($\mathsf{BEP}$)}. $\mathsf{BEP}$ quantifies how possible it is for users to escape from a filter bubble by comparing user behaviors with contrastive tendencies. Specifically, we design a contrastive simulation framework in which synthetic users exhibit either positive behavior—actively exploring new content—or negative behavior—reinforcing prior preferences. By comparing the exposure patterns generated by the same recommendation model for these two user types, we can decouple the effects of system bias from behavioral tendencies and provide a clearer diagnosis of bubble severity.

To implement this idea, we utilize large language model (LLM) agents \cite{yao2023react, wang2024survey}, which have recently emerged as powerful tools for simulating user interaction. These agents are capable of controlled planning, decision-making, and consistent behavior generation \cite{park2023generative, xie2024can}. Compared to traditional simulators or real-world datasets, LLM agents offer a distinct advantage: they allow us to precisely manipulate user goals and observe system responses under tightly controlled conditions. This makes them ideal for our framework, where accurate control of user behavior is key.

We evaluate $\mathsf{BEP}$ across multiple representative recommendation models and analyze the relationship between predictive accuracy and bubble severity. Our results not only validate $\mathsf{BEP}$'s ability to capture the trade-off between accurate preference modeling and information confinement, but also reveal a counter-intuitive finding: mild randomization in recommendation lists does \textit{not} effectively reduce filter bubbles. This insight highlights the need for more principled strategies in mitigating algorithmic confinement.

In summary, our contributions are as follows:
\begin{itemize}
    \item We introduce \textbf{Bubble Escape Potential ($\mathsf{BEP}$)}, a novel behavior-aware metric for measuring filter bubble severity by contrastive user behavioral intent.
    \item We present a contrastive simulation framework using LLM agents to systematically evaluate the influence of user behavior in recommendation scenarios.
    \item We conduct extensive empirical validation and uncover new insights into the complex trade-offs between personalization, diversity, and user freedom.
\end{itemize}

\section{Related Work}

\subsection{Filter Bubble}

The concept of the \textit{filter bubble} was first introduced and widely spread by Eli Pariser in 2011 \cite{pariser2011filter}. Later studies have focused on understanding its causes and finding ways to reduce its impact.

There are three main methods used to study the filter bubble: (1) static datasets \cite{sukiennik2024uncovering}, (2) simulating interactions between users and recommendation systems \cite{anwar2024filter}, and (3) mathematical modeling \cite{piao2023human}. \cite{sukiennik2024uncovering} finds that the filter bubble becomes stronger as item classification becomes more detailed. \cite{anwar2024filter} distinguishes the filter bubble from homogeneity, arguing that it can involve both high and low inter-user diversity. \cite{piao2023human} models the formation of the filter bubble using stochastic differential equations. What's more, \cite{ming2024modeling} develops an adaptive imitation process to further explore its causes and potential solutions. \cite{gao2023cirs} proposes a counterfactual interactive recommendation system that reduces the filter bubble by inferring information overexposure.  \cite{zhang2024practical} introduces a category-based retrieval method using a next-category prediction model to ease the filter bubble effect.

\subsection{LLM Agents for User Simulation in RS}

With the growing use of large language models (LLMs) in recommendation systems, researchers have started using LLM Agents as simulated users to enrich training data and explore system behaviors. RecAgent \cite{wang2023user} is the earliest framework to simulate users using LLM Agents in recommendation systems. Agent4Rec \cite{zhang2024generative} focuses on simulating real user behavior and modeling feedback from interactions. Recently, more user simulation frameworks have emerged, expanding the user characteristics and improving the alignment with real users \cite{zhang2025llm, cai2025agentic, liu2025agentcfplus}. Some of these frameworks \cite{wang2023user, zhang2024generative} have attempted to simulate the filter bubble effect. However, these efforts are still limited in depth.

\subsection{Recommendation Systems}


A recommendation system is an information filtering tool that delivers the most relevant content to a specific user, helping reduce information overload on modern internet platforms. Traditional collaborative filtering methods predict user preferences based on historical data \cite{sarwar2001item, koren2009matrix, he2017neural, wang2019adversarial, wang2021implicit}. Sequential recommendation focuses on leveraging the temporal order of user-item interactions \cite{hidasi2015session, kang2018self}. More recently, with the rapid development of deep learning \cite{han2024aucseg}, researchers have explored applying LLMs to recommendation systems \cite{sun2019bert4rec, li2023gpt4rec}.

\subsection{Diversified Recommendation}

Diversified recommendation has been a critical area of research in the field of recommender systems. It aims to balance relevance and diversity in the recommended items. MMR \cite{ziegler2005improving} is first to optimize both relevance and diversity by iteratively selecting items that maximize diversity. DPP \cite{kulesza2011determinantal} offers a probabilistic approach to model diversity by determinants. So far, diversified recommendations are extensively studied \cite{steck2018calibrated, zheng2021dgcn, liu2023personalized, yang2023dgrec, li2024contextual, coppolillo2024relevance}.

Although our work is based on diversity, there are fundamental differences. First, we consider the role of user behavior within the filter bubble. Second, $\mathsf{BEP}$ and diversity metrics differ in key aspects. We well explain it in Sec. \ref{sec:limit} and Sec. \ref{sec:bep}.

\section{Preliminary}

\subsection{Problem Definition}
In a recommendation system $R$, there are $N$ items denoted as $I=\{i_1,i_2,...,i_{N}\}$, categorized into $M$ classes. Each item $i_k$ belongs to a category $c_k$. For a user group $U$ in $R$, their interactions are observed over $T$ time periods, indexed by $t \in \{1,2,...,T\}$. At each time $t$, user $u \in U$ receives a list of recommended items from the system, represented as $L_{u,t}\subset I$. The user selects some of these items to interact with, resulting in the interaction set $S_{u,t}\subseteq L_{u,t}$. As $t$ increases, the diversity of information in $L_{u,t}$ is generally expected to decrease.

\subsection{Existing Metrics}

A common perspective in recent studies \cite{piao2023human, wang2023user, sukiennik2024uncovering, zhang2024generative, gao2023cirs} is that the reduction in diversity of recommended items over time, denoted as $diversity(L_{u,t})$, serves as an indicator of the filter bubble effect. Several metrics have been proposed to quantify $diversity(L_{u,t})$:
\begin{itemize}
    \item \textbf{Standardized information entropy} \cite{piao2023human}: 
    $\tilde s_{u,t}=\frac{s_{u,t}}{s_u^*}$
    , where $s_u^*$ is a user-specific normalization term, and $s_{u,t}=-\sum_{c=1}^{M}f_{u,t}^clogf_{u,t}^c$. Here, $f_{u,t}^c$ denotes the proportion of category $c$ in $L_{u,t}$.
    \item \textbf{Category coverage rate} $C_{u,t}$ \cite{sukiennik2024uncovering}: 
    $C_{u,t}=\frac{cate(L_{u,t})}{cate(I)}$
    , where $cate(L)$ is the number of distinct categories in the item set $L$.
    \item \textbf{Top-1 genre percentage} $P_{top-1}$ \cite{zhang2024generative}: the average proportion of the most frequent genre among the recommended movies.
\end{itemize}

These metrics are generally consistent in how they capture diversity. Most studies rely on simulation or statistical analysis to track how $diversity(L_{u,t})$ evolves over time \cite{piao2023human, wang2023user, sukiennik2024uncovering, zhang2024generative, zhang2024practical}. By comparing the trend of diversity decline across different recommendation algorithms or user groups, researchers aim to uncover the underlying causes of filter bubbles.

\subsection{Limitations of Existing Metrics} \label{sec:limit}

Although these methods provide some insights into the severity of filter bubbles, they generally have a substantial limitation. Specifically, the decline in diversity may result from various factors beyond the filter bubble, including the users’ own preferences and behavior. \textbf{However, they do not clearly distinguish between the influence of filter bubbles and the natural outcome of preference modeling, which may create a misleading correlation between accuracy and filter bubble severity.}

\begin{figure}
    \centering
    \includegraphics[width=0.7\linewidth]{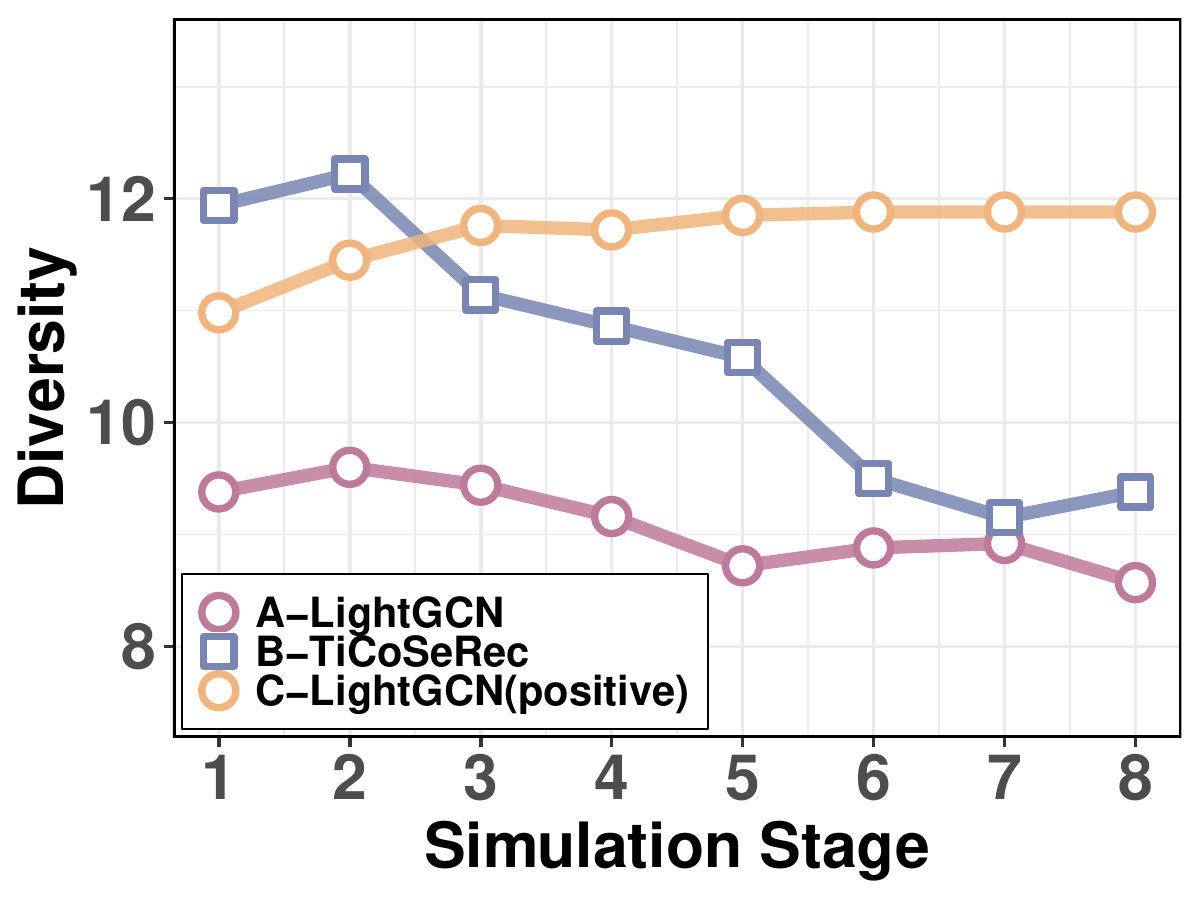}
    \caption{Change in diversity under different settings: A-LightGCN, B-TiCoSeRec, and C-LightGCN(positive) in \textit{ml-1m}. Users' positive behaviors in LightGCN surpasses the gap between different recommendation systems.}
    \label{fig:case_study}
\end{figure}

To better understand these concerns, we conduct a case study using simulated users. We present three line graphs in Figure \ref{fig:case_study}. It shows the decline in diversity under different settings: Line A, B, and C. Line A uses LightGCN \cite{he2020lightgcn}, line B replaces LightGCN with TiCoSeRec \cite{dang2023ticoserec}, and line C retains LightGCN but modifies user behaviors to be more positive (details discussed later). We find that the diversity in Group C decreases more slowly than in Group B, suggesting that user behavior significantly affects diversity trends—sometimes even more than the choice of recommendation algorithm.

Hence, it is necessary to explore a more reasonable measure for filter bubbles, which can better decouple filter bubbles from the other factors, which we will elaborate on later.

\section{Method}

\begin{figure*}[htb]
  \centering
  \begin{subfigure}[b]{0.23\linewidth}
    \includegraphics[width=\linewidth]{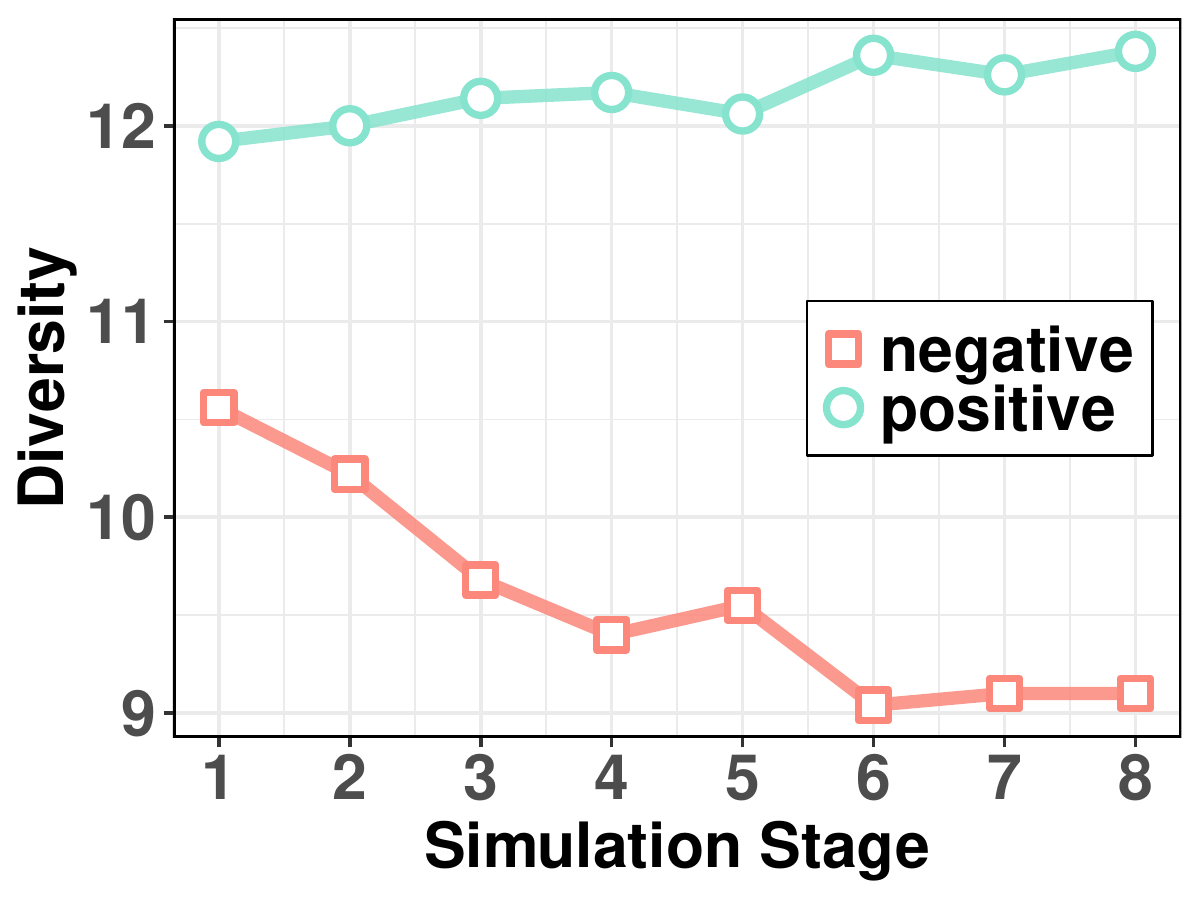}
    \caption{}
  \end{subfigure}
  \hfill
  \begin{subfigure}[b]{0.23\linewidth}
    \includegraphics[width=\linewidth]{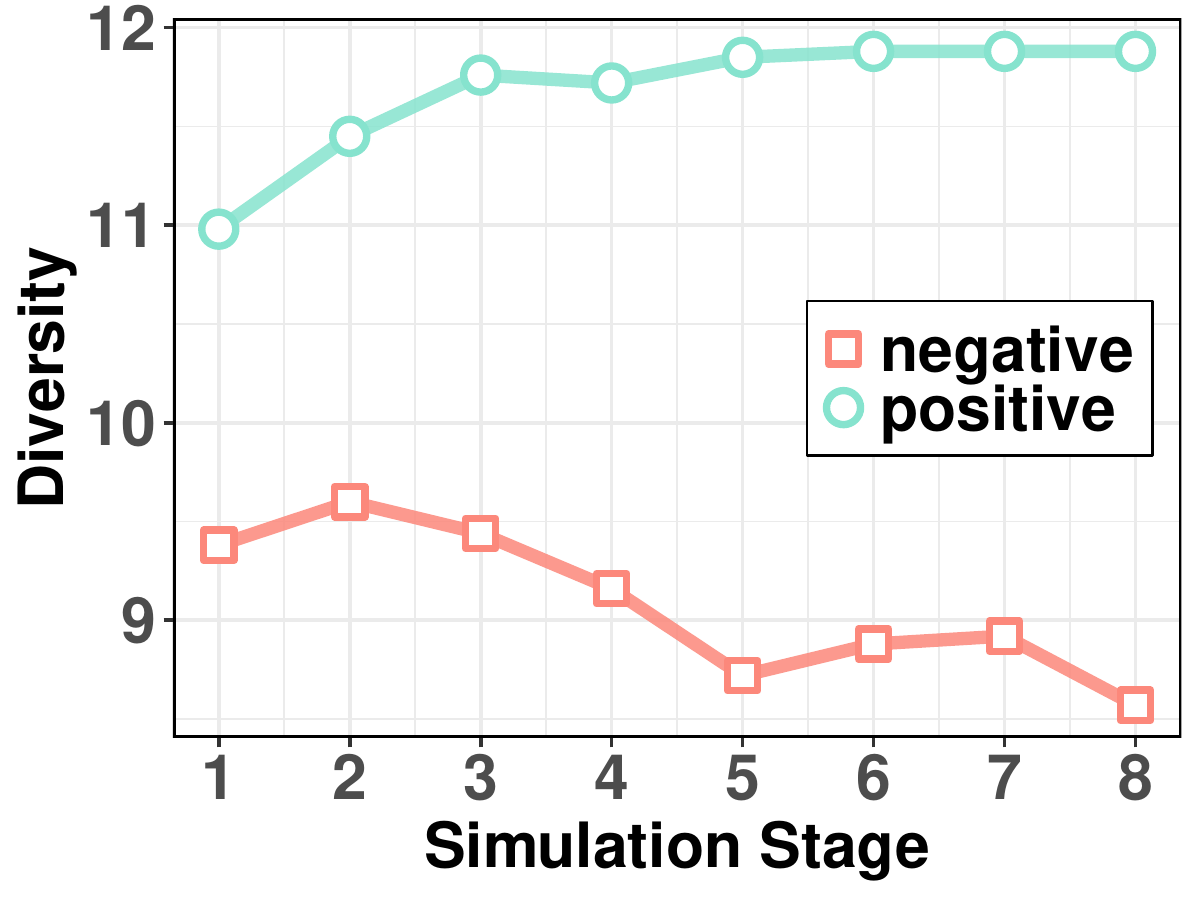}
    \caption{}
  \end{subfigure}
  \hfill
  \begin{subfigure}[b]{0.23\linewidth}
    \includegraphics[width=\linewidth]{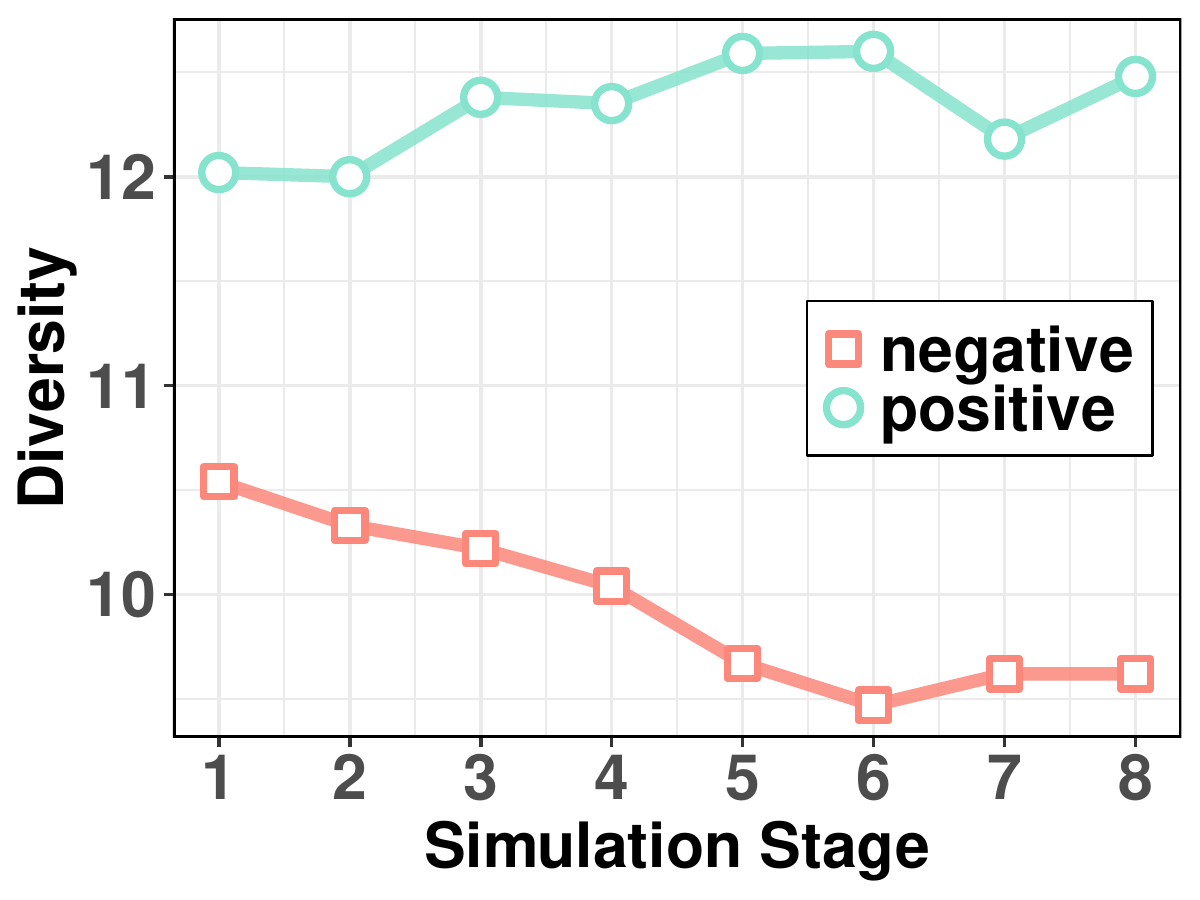}
    \caption{}
  \end{subfigure}
  \hfill
  \begin{subfigure}[b]{0.23\linewidth}
    \includegraphics[width=\linewidth]{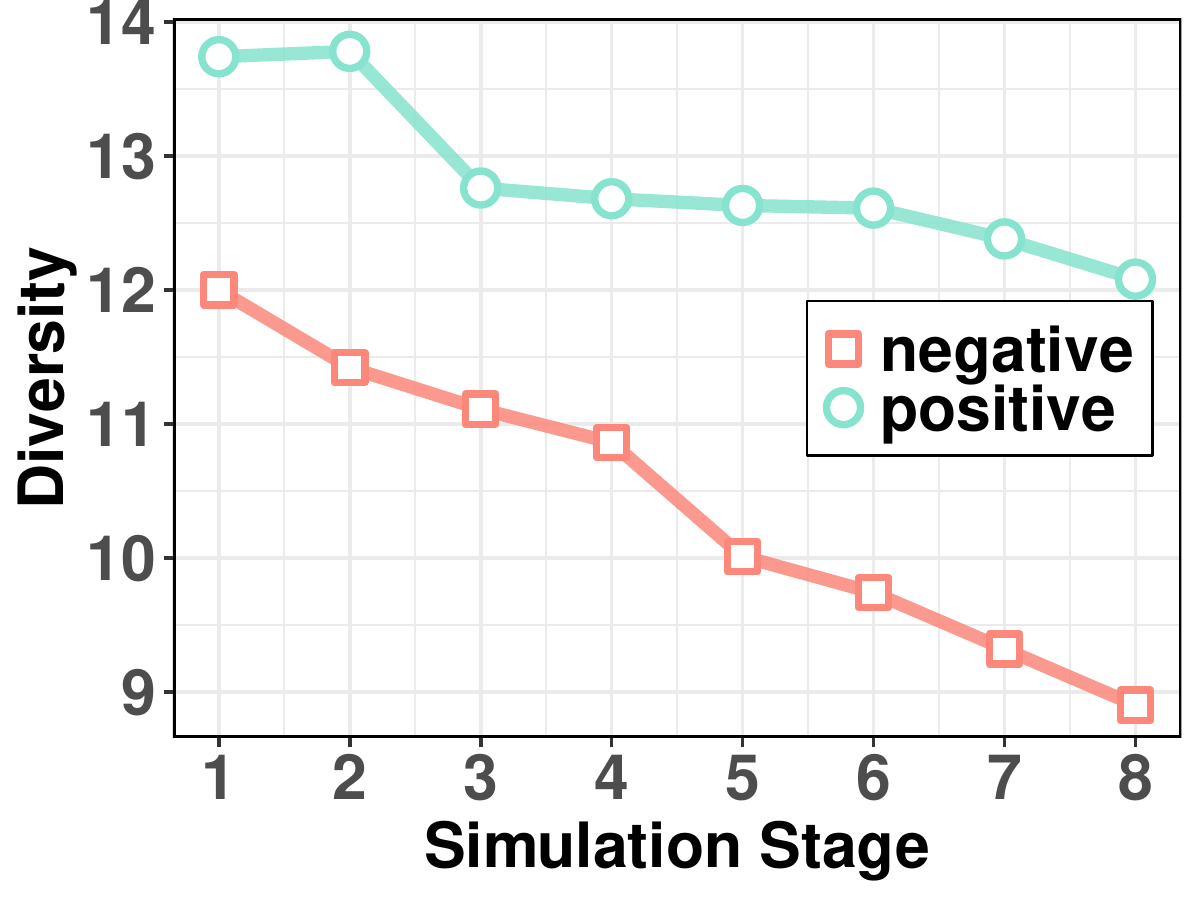}
    \caption{}
  \end{subfigure}
  \caption{The variation in the diversity of recommendations received by users under different behavioral patterns changes over time as the simulation rounds increase in different recommendations in \textit{ml-1m}. Blue lines represent positive users, while red lines represent negative users.  Each of these pictures represents a different recommendation system: (a) MF. (b) LightGCN. (c) DiffRec. (d) TiCoSeRec.}
  \label{fig:trend}
\end{figure*}

The overall process of our approach is illustrated in Figure~\ref{fig:procedure}. In this section, we describe the workflow in three parts. First, we explain how user behaviors are modeled. Second, we describe how interaction data is collected through simulation. Last, we illustrate how our new metric, \textbf{Bubble Escape Potential ($\mathsf{BEP}$)}, is calculated.

\subsection{Behaviors of Users}



We conduct two separate user simulations to study the impact of filter bubbles: one with users assigned positive behavior, and the other with users assigned negative behavior. In both simulations, users are modeled using agents powered by large language models (LLMs). The general format of the user simulation prompts follows the settings introduced in \cite{zhang2024generative}.

Each simulated user is defined by two types of characteristics: social traits and unique tastes. Unique tastes are extracted and summarized by LLMs from real users’ browsing histories. Social traits are derived from real-world data and include three aspects: a) \textit{Activity}: the frequency and range of a user’s interactions. b) \textit{Conformity}: the extent to which a user’s ratings align with the average item ratings. c) \textit{Diversity}: a user’s tendency to engage with different types or categories of items.

To simulate positive behavior, we assign users a prompt that encourages them to actively escape filter bubbles. We also set their activity and diversity levels to the highest values. The prompt used in this simulation is shown in the blue box. In the prompt box, the specific type of \textit{[item]} varies depending on the type of the item.

To simulate negative behavior, we assign users a prompt that encourages reliance on the recommendation system. We also set their activity and diversity levels to the lowest values. The corresponding prompt is shown in the red box.

\subsection{User Simulation}

In the user simulation, a group of users $U$ interacts with the recommendation system $R$ for $T$ rounds. During the simulation, the system continuously collects users' interaction data $\{S_{u,t}\}$ and updates its model through re-training.

\begin{tcolorbox}[
    float,
    label=box:positive,
    title=Prompt of Positive Behavior,
    coltitle=black,
    colback=white,
    colframe=gray,
    colbacktitle=lightblue,
    fonttitle=\bfseries,
    rounded corners,
    enhanced,
]
\textit{You will actively broaden your [item] horizons. You will actively watch [item] that do not match your [item] tastes. Specifically, when you see the recommended list, while you are watching [item] that align with your tastes, you will also watch a few [item] that do not match your tastes.}
\end{tcolorbox}
\begin{tcolorbox}[
    float,
    title=Prompt of Negative Behavior,
    coltitle=black,
    colback=white,
    colframe=gray,
    colbacktitle=lightpink,
    fonttitle=\bfseries,
    rounded corners,
    enhanced
]
\textit{You rely heavily on the recommendation system. You believe that [item] recommended by the recommendation system will match your own [item] tastes. Specifically, when you see the recommended list, you will watch a few top-ranked [item] first, and then watch other [item] based solely on your own taste.}
\end{tcolorbox}
At the start of the simulation, we perform a cold-start initialization using an interaction set extracted from an official dataset, denoted as $A_0$. In each round $t$, the recommendation model is trained on $A_{t-1}$ to produce a new model $R'_t$. Then, each user $u$ receives a recommendation list $L_{t,u}$ and selects a set of items $S_{t,u}$ to interact with based on predefined behaviors. All interactions in round $t$ are then merged into $A_{t-1}$ to form a new interaction set $A_t$. After completing all $T$ rounds, we collect all recommendation lists $\{L_{u,t}\}$ and use the previously described method to compute the \textbf{Bubble Escape Potential}.



\subsection{Bubble Escape Potential} \label{sec:bep}

Given a recommendation system $R$, we define its corresponding \textbf{Bubble Escape Potential}~as $\mathsf{BEP}(R)$, which quantifies the probability of users to escape from the filter bubble induced by $R$. To estimate $\mathsf{BEP}(R)$, we compare the diversity of items recommended to two groups of simulated users: one exhibiting positive behaviors and the other exhibiting negative behaviors. 

We begin by assigning all agents in user group $U$ with positive behaviors and simulating their interactions with the recommendation system over $T$ consecutive rounds. The full recommendation history for user $u$ is:
\begin{equation}
L_u = \{L_{u,1}, L_{u,2}, ..., L_{u,T}\}
\end{equation}

\noindent Taking a further step, we define the diversity of recommendations for user $u$ at round $t$ as the number of distinct categories in the list:
\begin{equation}
    D_{u,t} = \left| \left\{\, c(i) \mid i \in L_{u,t} \,\right\} \right|
\end{equation}
where $c:\mathcal{I}\rightarrow \mathcal{C}$ is a mapping from items to their categories and $\mathcal{C}$ is the set of all possible categories. This gives a sequence of diversity values for each user:
\begin{equation}
    D_u = \{D_{u,1}, D_{u,2}, ..., D_{u,T}\}
\end{equation}

\noindent We then reassign the same users with negative behaviors and repeat the simulation for another $T$ rounds, collecting their corresponding diversity values $D'_u$ in the same way.

For each round $t$, the estimated escape potential $\widehat{\mathsf{BEP}}_{R,t}$ is defined as the ratio between the total diversity of the positive-behavior users and that of the negative-behavior users:
\begin{equation}
\widehat{\mathsf{BEP}}_t(R) = \frac{\sum_{u \in U} D_{u,t}}{\sum_{u \in U} D'_{u,t}}
\end{equation}

Finally, the overall escape potential for system $R$ is calculated by averaging over all rounds:
\begin{equation}
\widehat{\mathsf{BEP}}(R) = \frac{1}{T} \sum_{t=1}^{T} \widehat{\mathsf{BEP}}_{R,t}.
\end{equation}

Based on the previous description of user behaviors, it is expected to have $\mathsf{BEP}(R) \geq 1$. Moreover, it can be known that an important property exists that \textbf{the smaller the value of $\mathsf{BEP}(R)$, the more severe the filter bubble in recommendation system $R$}.

In a nutshell, the advantages of our metric can be listed as follows:
\begin{itemize}
    \item By contrasting users with different behavioral intents, $\mathsf{BEP}$ decouples the influence of user preference modeling from system-induced confinement.
    \item It enables precise diagnosis of filter bubble without relying on assumptions about user intent or model internals.
    \item Our experiments using $\mathsf{BEP}$ are the first to quantitatively validate the inherent tension between accurate preference modeling and the emergence of filter bubbles.
    \item When the diversity of information received by positive and negative users increases by the same proportion simultaneously, $\mathsf{BEP}$ remains unchanged. This property makes it robust to uniform diversification strategies.
\end{itemize}

\section{Experiments}

\begin{figure*}[htb]
  \centering
  \begin{subfigure}[b]{0.23\linewidth}
    \includegraphics[width=\linewidth]{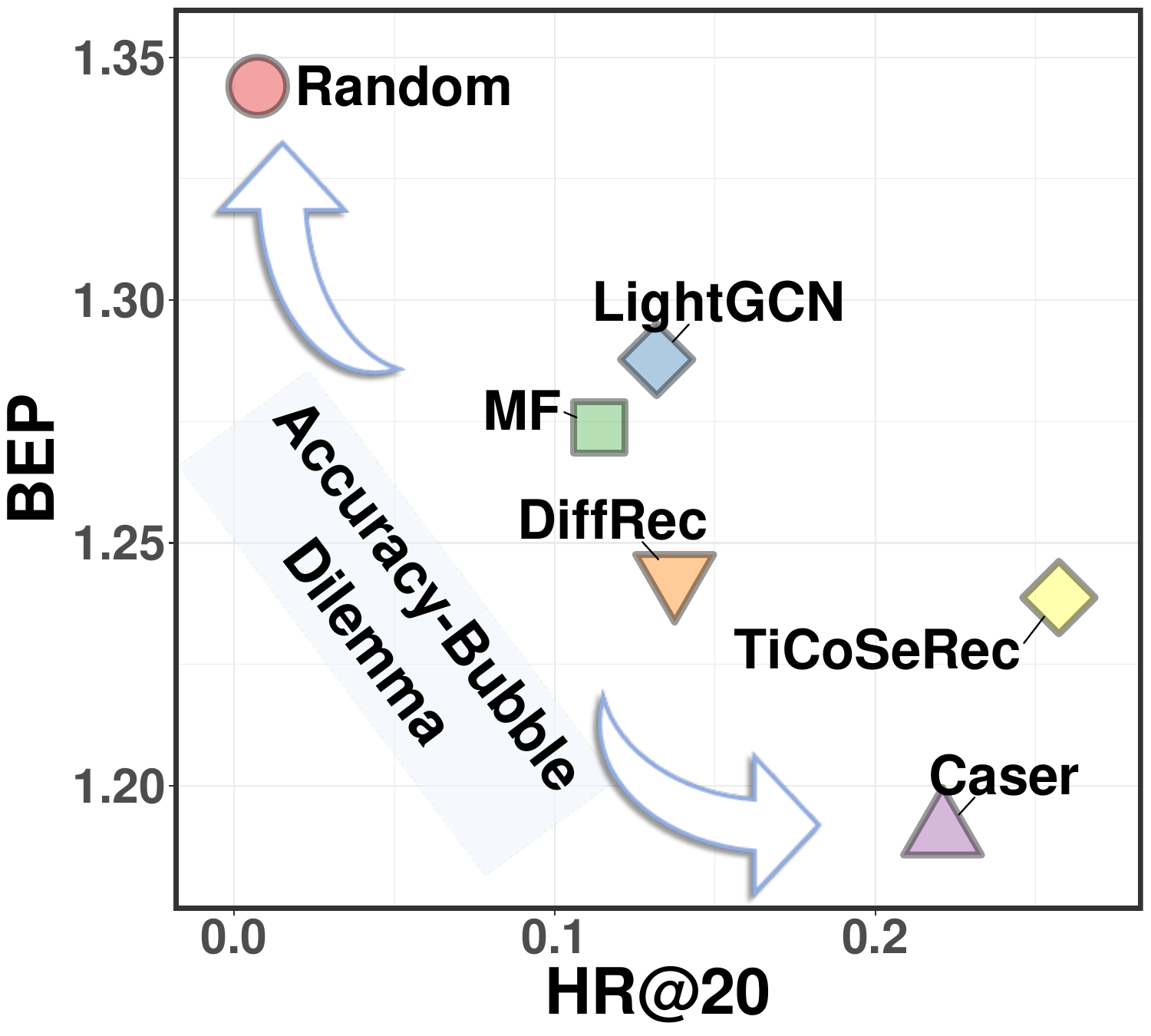}
    \caption{}
  \end{subfigure}
  \hfill
  \begin{subfigure}[b]{0.23\linewidth}
    \includegraphics[width=\linewidth]{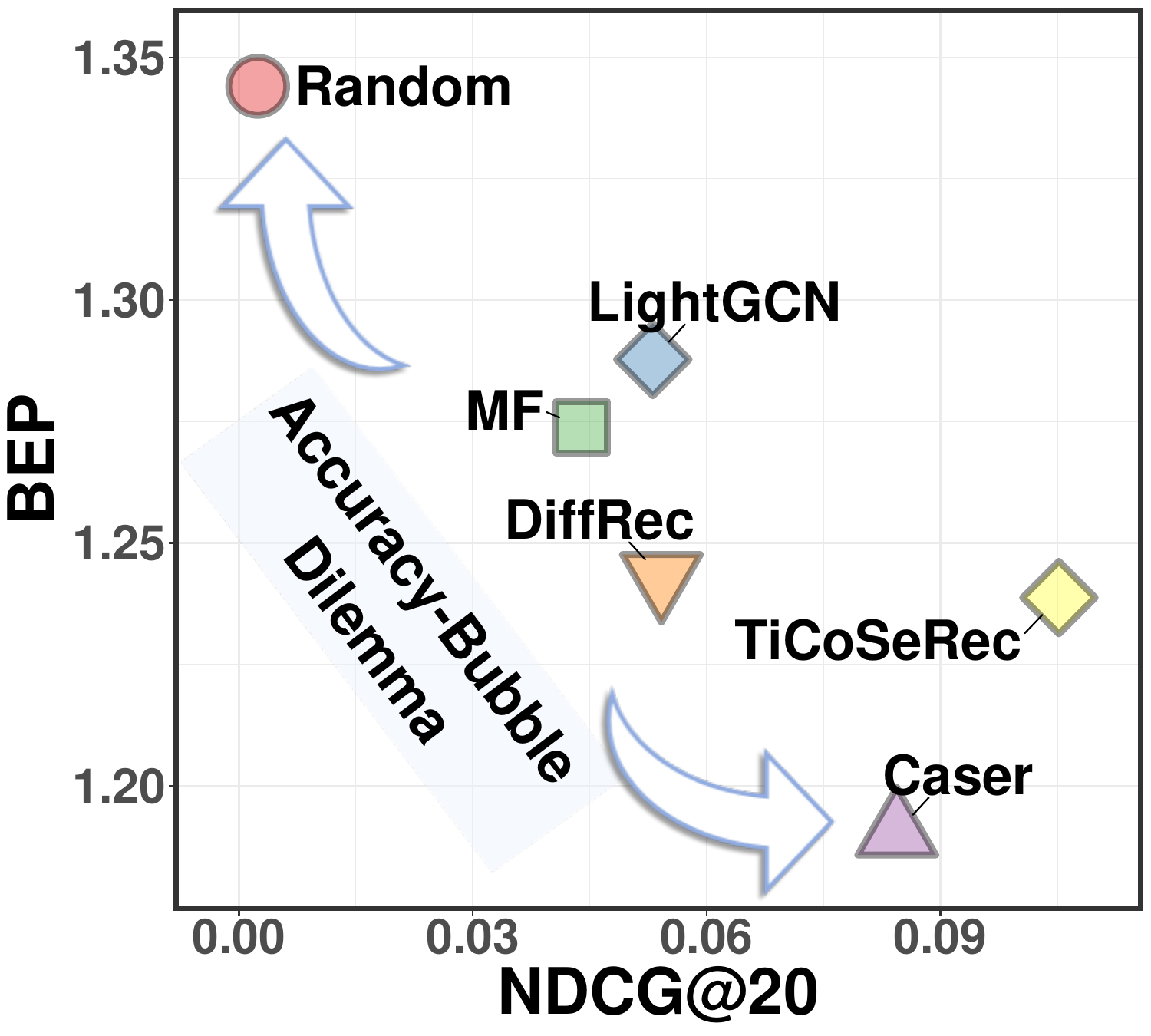}
    \caption{}
  \end{subfigure}
  \hfill
  \begin{subfigure}[b]{0.23\linewidth}
    \includegraphics[width=\linewidth]{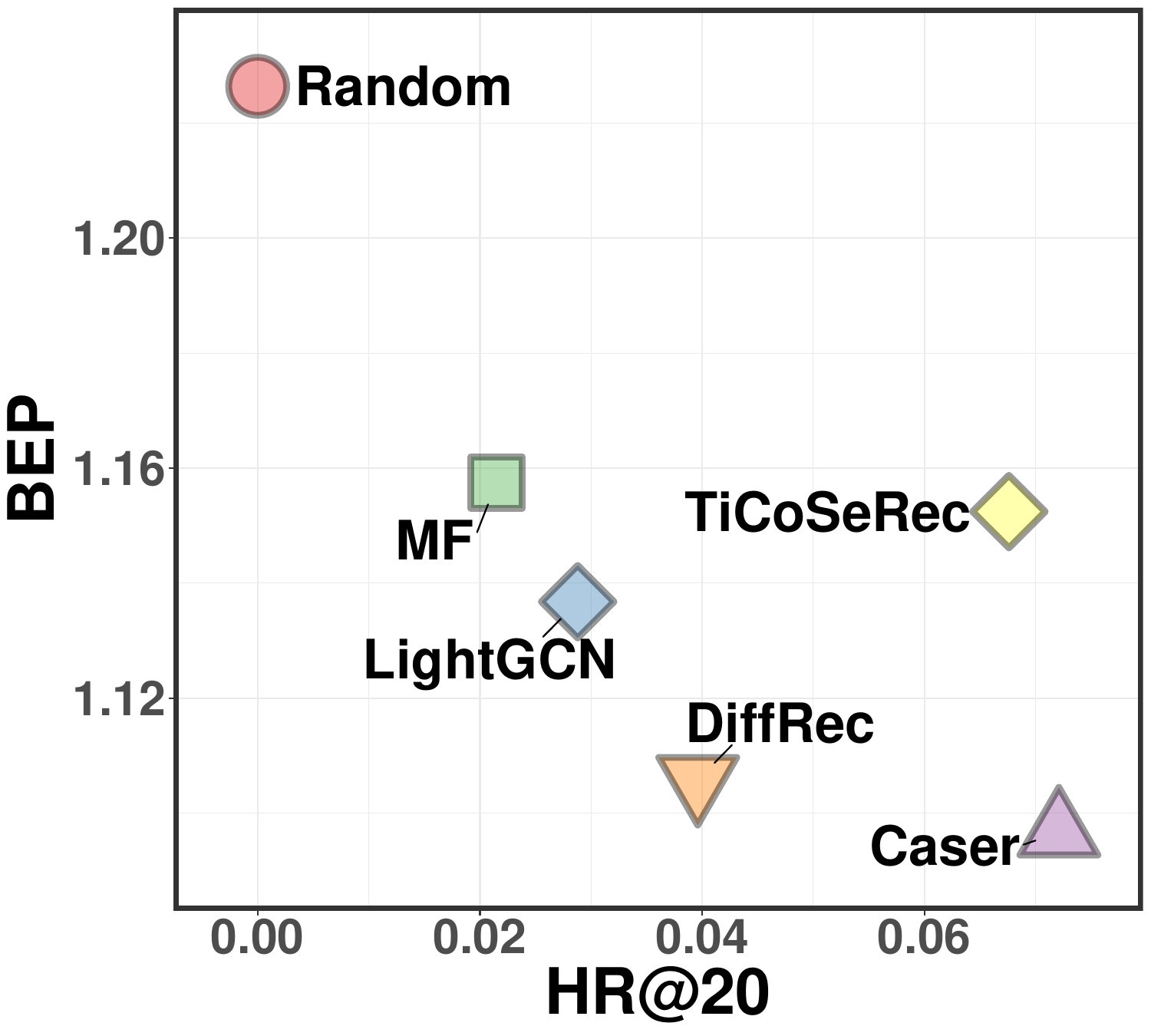}
    \caption{}
  \end{subfigure}
  \hfill
  \begin{subfigure}[b]{0.23\linewidth}
    \includegraphics[width=\linewidth]{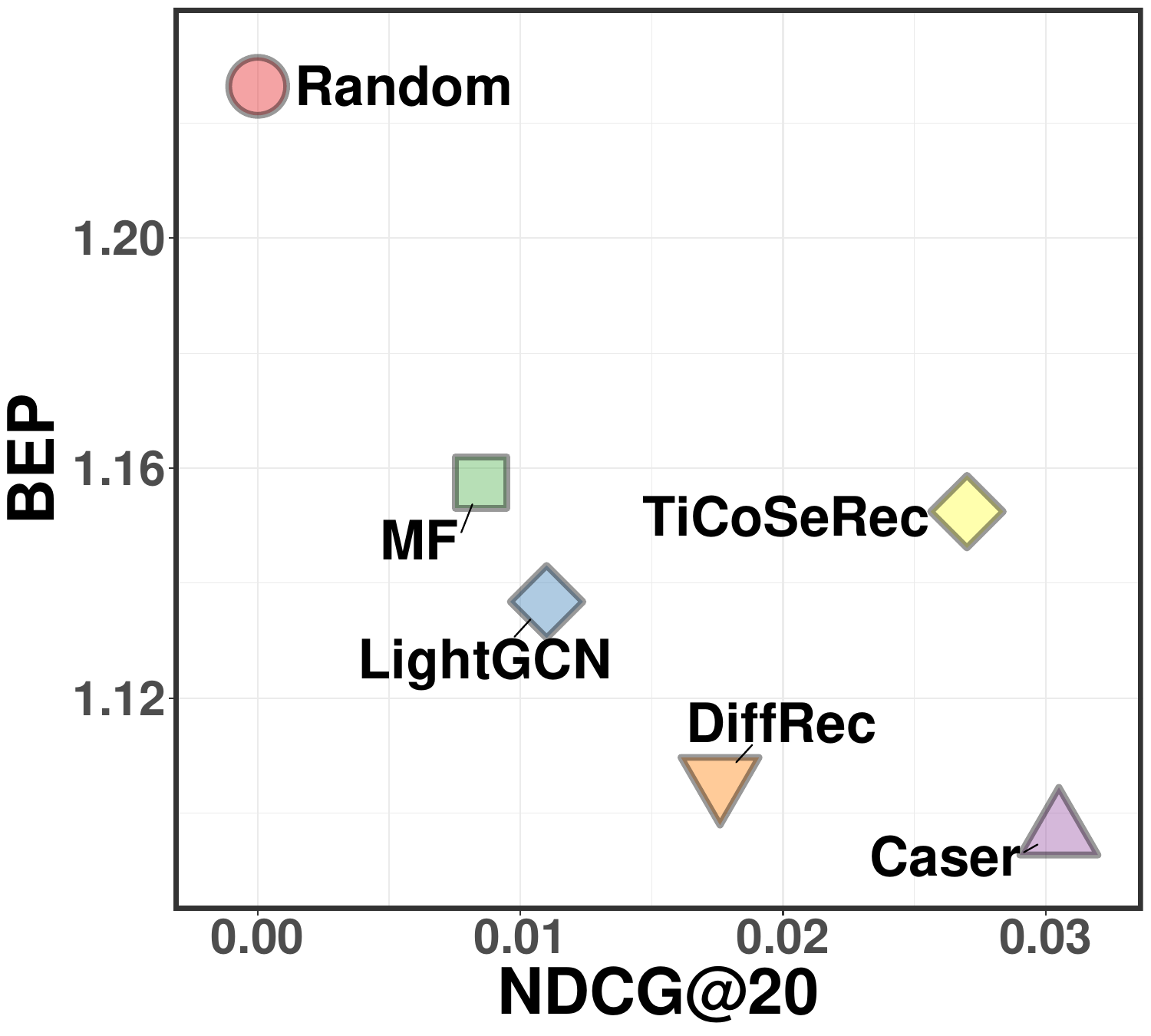}
    \caption{}
  \end{subfigure}
  \caption{The accuracy and the Bubble Escape Potential ($\mathsf{BEP}$) corresponding to different recommendation systems under \textit{ml-1m} and \textit{Amazon-Books}: (a) HR@20 vs. $\mathsf{BEP}$ in \textit{ml-1m}. (b) NDCG@20 vs. $\mathsf{BEP}$ in \textit{ml-1m}. (a) HR@20 vs. $\mathsf{BEP}$ in \textit{Amazon-Books}. (b) NDCG@20 vs. $\mathsf{BEP}$ in \textit{Amazon-Books}. This set of results reveals the \textit{Accuracy-Bubble Dilemma}.}
  \label{fig:tradeoff_4images}
\end{figure*}

\begin{figure}[htb]
  \centering
  \begin{subfigure}[b]{0.48\linewidth}
    \includegraphics[width=\linewidth]{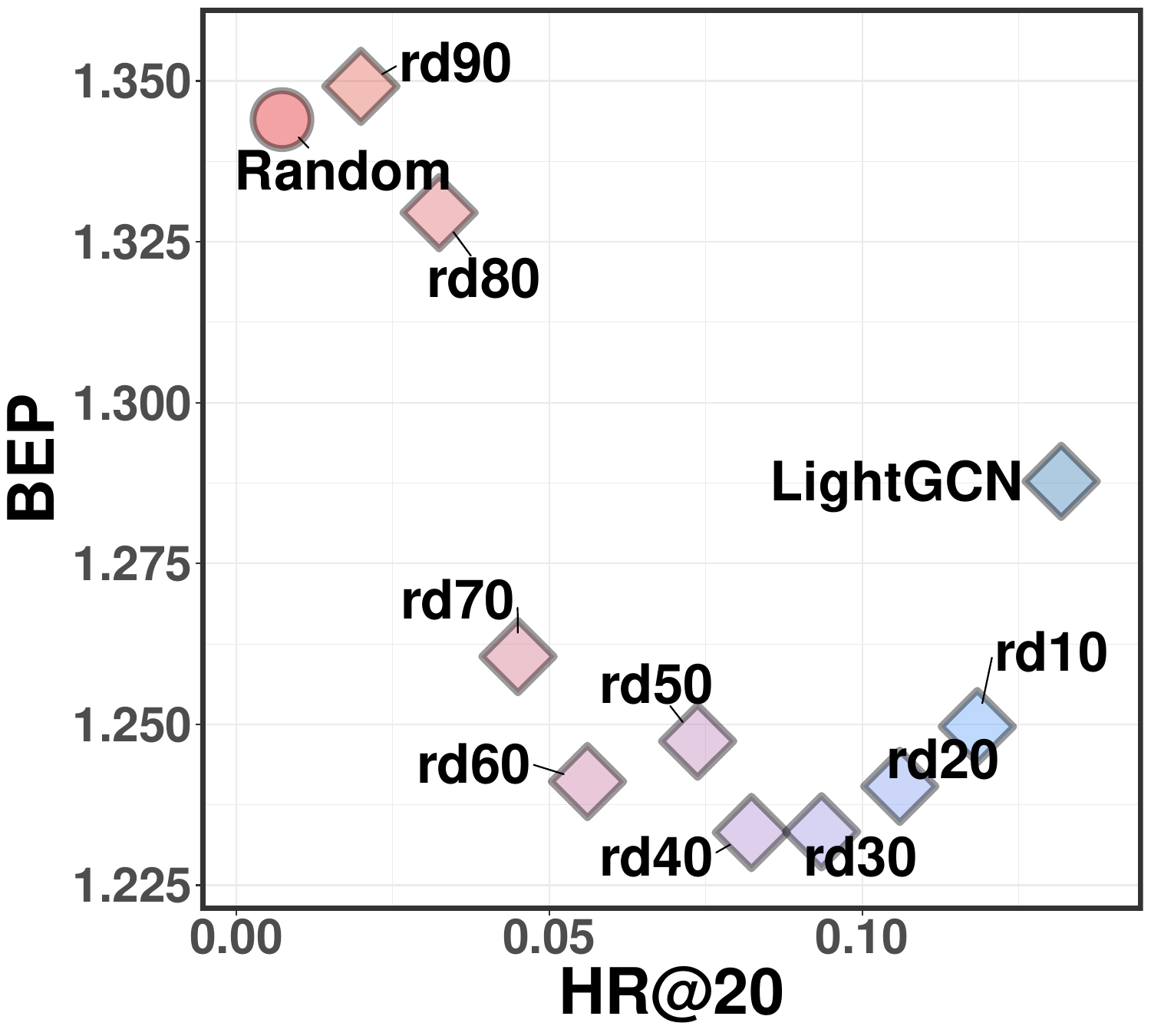}
    \caption{}
  \end{subfigure}
  \hfill
  \begin{subfigure}[b]{0.48\linewidth}
    \includegraphics[width=\linewidth]{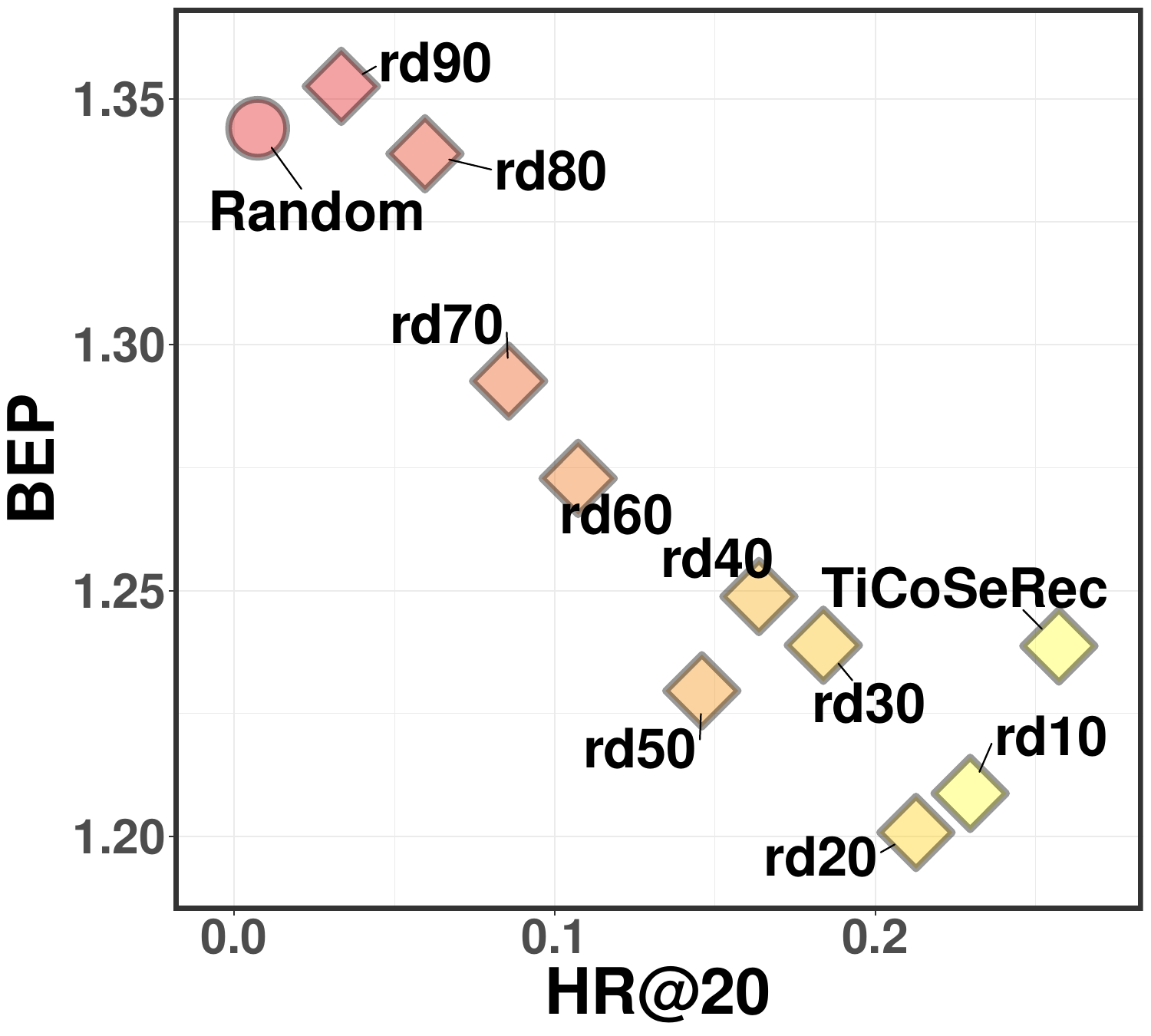}
    \caption{}
  \end{subfigure}
  \caption{The accuracy and the Bubble Escape Potential ($\mathsf{BEP}$) corresponding to different recommendation systems after adding randomization under \textit{ml-1m}: (a) LightGCN. (b) TiCoSeRec.}
  \label{fig:tradeoff_2images}
\end{figure}


In this section, we present the experimental setup and results. The experiments aim to answer the following research questions:

\begin{itemize}
    \item \textbf{(RQ1)} To what extent can the simulated users approximate the real users?
    \item \textbf{(RQ2)} Do the positive and negative behavior settings significantly influence user actions?
    \item \textbf{(RQ3)} Is it sufficient to set two types of behaviors?
    \item \textbf{(RQ4)} What is the relationship between accuracy and $\mathsf{BEP}$ in different recommendation systems?
    \item \textbf{(RQ5)} Can introducing randomness into recommendation strategies help balance accuracy and $\mathsf{BEP}$?
    \item \textbf{(RQ6)} How do user groups with different characteristics vary in their potential to escape filter bubbles?
\end{itemize}

\subsection{Datasets}

We conduct our experiments on two real-world datasets:
\begin{itemize}
    \item \textbf{MovieLens-1M} \cite{harper2015movielens}: a widely used benchmark dataset of movies in collaborative filtering. We select the 1M version as \textit{ml-1m}.
    \item \textbf{Amazon-Books} \cite{ni-etal-2019-justifying}: a large-scale real-world dataset collected from Amazon’s book category, comprising user reviews and ratings.
\end{itemize}





\subsection{Baselines \& Metrics}

To evaluate our proposed metric comprehensively, we select several representative and recent recommendation models: Random recommendation, BPR-MF \cite{rendle2012bpr}, LightGCN \cite{he2020lightgcn}, Caser \cite{tang2018personalized}, DiffRec \cite{wang2023diffusion}, and TiCoSeRec \cite{dang2023ticoserec}. We evaluate recommendation accuracy by HR@k, NDCG@k, and MAP. To measure the severity of filter bubbles, we use our proposed metric, Bubble Escape Potential ($\mathsf{BEP}$).

\subsection{Implementation Details}
We adopt the \textit{leave-one-out strategy} \cite{wang2021counterfactual, liu2021contrastive, dang2023ticoserec} to prepare the test data. For each user's behavior sequence, the last interacted item is used as the test set, while the remaining interactions form the training set. For each user-item pair $(u, i_u)$ in the test set, we record the position of item $i_u$ in the recommendation list $L_u$ generated by the system as $p_u$.

For user simulation, following Agent4Rec \cite{zhang2024generative}, we first select 1000 users with frequent interactions to form the cold-start dataset. Then, we randomly select 200 users from this group and infer their social characteristics and personal preferences based on their historical behaviors. The global parameters used in the experiment are set as $N=200, T=8$. To guarantee reproduction, we use Qwen2.5-14B-Instruct-1M \cite{yang2025qwen2} as the LLM supporting the user group. For the recommendation models, we use the official implementations and retain their default settings.



\subsection{Results}


\begin{figure*}[htb]
  \centering
  \begin{subfigure}[t]{0.16\linewidth}
    \raisebox{2.7ex}{\includegraphics[width=\linewidth]{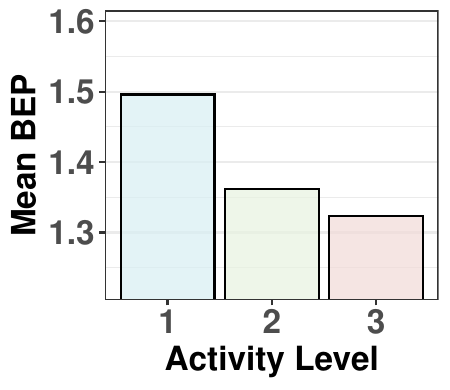}}
    \caption{}
  \end{subfigure}
  \hfill
  \begin{subfigure}[t]{0.16\linewidth}
    \raisebox{2.7ex}{\includegraphics[width=\linewidth]{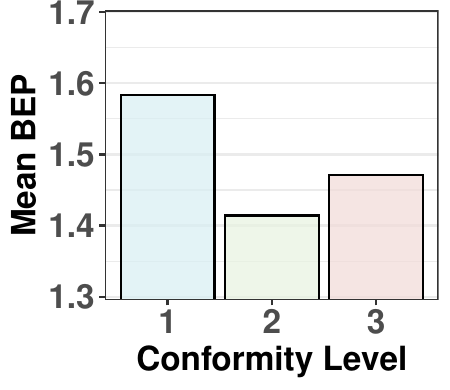}}
    \caption{}
  \end{subfigure}
  \hfill
  \begin{subfigure}[t]{0.16\linewidth}
    \raisebox{2.7ex}{\includegraphics[width=\linewidth]{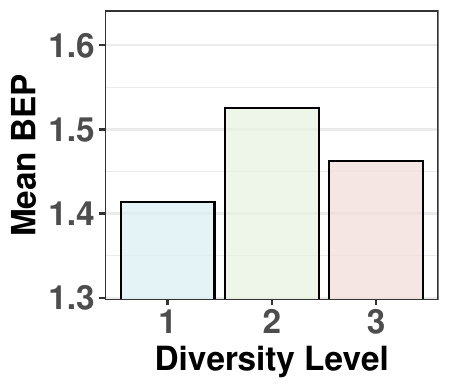}}
    \caption{}
  \end{subfigure}
  \hfill
  \begin{subfigure}[t]{0.5\linewidth}
    \includegraphics[width=\linewidth]{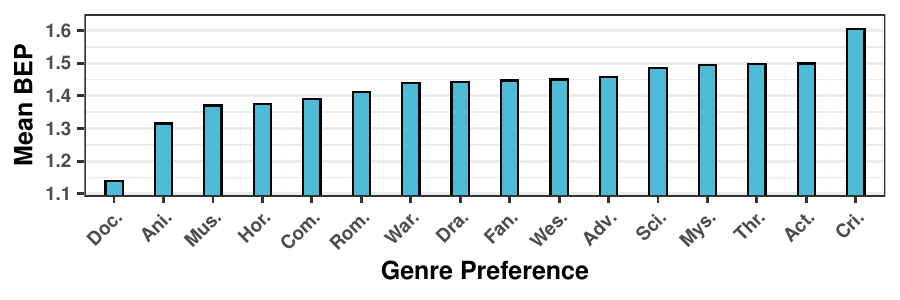}
    \caption{}
  \end{subfigure}
    \caption{
        \textbf{Comparison of Mean Bubble Escape Potential ($\mathsf{BEP}$) across user characteristics.}
        (a-c): Three intrinsic user profiles—(a) \textit{Activity Level}, (b) \textit{Conformity Level}, and (c) \textit{Diversity Level}. Users with lower conformity and higher diversity tend to have higher $\mathsf{BEP}$, suggesting that if they choose positive actions, it will lead to a much broader expansion of the range of information they can accept.
        (d): Mean $\mathsf{BEP}$ grouped by \textit{Genre Preference}. 
    }
    \label{fig:bep_group_analysis}
\end{figure*}




\noindent \textbf{Reality of simulated users (RQ1).} In Table~\ref{tab:acc} and Table~\ref{tab:rating}, we adopt the same method as \cite{zhang2024generative} to test the alignment between the simulated users and the real users, including the interaction accuracy and the distribution of ratings. Based on these tables, we observe the following:
\begin{itemize}
    \item The alignment accuracy of the simulated users is around 70\%. This validates the reality of users.
    \item The distribution of simulated users' ratings is similar to that of real users ($D_{KL}(P||Q)\approx0.125$). This indicates that simulated user ratings are similar to real users'.
\end{itemize}
\begin{table}[tb]
\centering
\begin{tabular}{ccccc}
\toprule
\textbf{Behavior} & Prediction & Recall & Accuracy & F1 Score \\
\midrule
positive & 0.67 & 0.76 & 0.68 & 0.70 \\
negative & 0.72 & 0.58 & 0.68 & 0.63 \\
\bottomrule
\end{tabular}
\caption{The degree of alignment between the preferences of simulated users and real users for different behaviors.}
\label{tab:acc}
\end{table}
\begin{table}[tb]
\centering
\begin{tabular}{ccc}
\toprule
Rating & Ratio of Agent ($P$) & Ratio of Users in \textit{ml-1m} ($Q$) \\
\midrule
1 & 0.001 & 0.056 \\
2 & 0.054 & 0.108 \\
3 & 0.164 & 0.261 \\
4 & 0.436 & 0.349 \\
5 & 0.345 & 0.226 \\
\bottomrule
\end{tabular}
\caption{Rating distribution comparison between agent-simulated users ($P$) and real users in the \textit{ml-1m} dataset ($Q$).}
\label{tab:rating}
\end{table}
\noindent \textbf{Distinction of behaviors (RQ2).} Figure~\ref{fig:trend} illustrates how the diversity of recommended items evolves across simulation stages for both positive and negative users under four recommendation systems in \textit{ml-1m} dataset. We will present the precise results of these figures in the supplementary material. From these results, we observe the following: 
\begin{itemize}
    \item Under the influence of negative behaviors, the diversity of information received by users significantly decreases.
    \item Under the influence of positive behaviors, the diversity of information received by users actually increases (Figure \ref{fig:trend}(a-c)), and in some cases it is suppressed (Figure \ref{fig:trend}(d)).
    \item Regardless of user type, the diversity stabilizes in later stages with only minor fluctuations.
    \item Across all methods, positive users consistently receive more diverse recommendations than negative users. 
\end{itemize}

\noindent \textbf{Test of weakened behavior (RQ3).} For calculating $\mathsf{BEP}$, just two behaviors are sufficient. Nevertheless, we still design two weakened behaviors (weakly positive and weakly negative), and calculate the $\mathsf{BEP}$ in the same form on \textit{ml-1m}. The results of random and TiCoSeRec are in Table \ref{tab:weak}. We find that their $\mathsf{BEP}$ decreases, and $\mathsf{BEP}$ of Random is still greater than that of TiCoSeRec. This is in line with our expectations.

\begin{table}[tb]
\centering
\begin{tabular}{lccc}
\toprule
\textbf{Models} & HR@20 & $\mathsf{BEP}$ & \textbf{$\mathsf{BEP}$-weak} \\
\midrule
Random & 0.001 & 1.35 & 1.20 \\
TiCoSeRec & 0.257 & 1.24 & 1.05 \\
\bottomrule
\end{tabular}
\caption{Comparison of recommendation performance and Bubble Escape Potential ($\mathsf{BEP}$) using original and weakened behavior (denoted as $\mathsf{BEP}$-weak) on the \textit{ml-1m} dataset.
}
\label{tab:weak}
\end{table}

\noindent \textbf{Correlations between $\mathsf{BEP}$ and accuracy (RQ4).} Figure~\ref{fig:tradeoff_4images} demonstrates the trade-off between recommendation accuracy and the severity of filter bubbles, as quantified by Bubble Escape Potential, in \textit{ml-1m} and \textit{Amazon-Books}. Notably, a lower $\mathsf{BEP}$ indicates a more severe filtering effect, meaning that users are more deeply trapped in their personalized content loops. From the results, we observe the following:
\begin{itemize}
    \item \textbf{There is a dilemma between accuracy and filter bubble.} As shown in Figure~\ref{fig:tradeoff_4images}, we observe negative correlation between accuracy and $\mathsf{BEP}$: models that achieve higher HR@20 or NDCG@20 like Caser and TiCoSeRec tend to produce lower $\mathsf{BEP}$, implying stronger filter bubble effects. In contrast, less accurate models like Random or DiffRec yield higher $\mathsf{BEP}$, suggesting weaker behavioral reinforcement and broader exposure.
    \item Moreover, a clear structural distinction is observed between different model types. Non-sequential models (MF, DiffRec, LightGCN) are clustered in the upper-left regions of the plots, while sequential models (Caser, TiCoSeRec) are in the lower-right. This indicates that sequential models generally provide higher accuracy but at the cost of more severe filter bubble formation. One possible explanation is that sequential models place greater emphasis on recent user behavior sequences, potentially narrowing the diversity of exposed content and overlooking long-term or global user preferences.
\end{itemize}

\noindent \textbf{Impact of introducing randomness on recommendation systems (RQ5).}  
To investigate the effect of controlled noise on filter bubble severity, we modify the output of two representative models, LightGCN and TiCoSeRec, by randomly replacing a portion $k$ ($k = 10\%, 20\%, 30\%, ... , 90\%$) of their recommendation lists with items sampled randomly. Figure~\ref{fig:tradeoff_2images}(a) \& (b) track the models' trajectories on the accuracy versus $\mathsf{BEP}$ plots as the level of randomness increases. The results reveal the following observations:
\begin{itemize}
    \item The impact of randomness is non-monotonic. As randomness $k$ increases to 30\%, $\mathsf{BEP}$ drops slightly. Surprisingly, the $\mathsf{BEP}$ reaches its \textit{lowest} point around $k=30\%$, implying that small-scale randomization may inadvertently reinforce personalization biases. However, as randomness continues to increase, $\mathsf{BEP}$ begins to rise, returning to its original level near $k=70\%$ and eventually surpassing it. At $k=80\%\sim90\%$, the models approach the performance of a fully random recommender. These suggest that introducing randomness does not help recommendation systems better balance the prevention of filter bubble and accuracy.
\end{itemize}


\noindent \textbf{The Bubble Escape Potential of different user groups (RQ6).}
Based on $\mathsf{BEP}$, we analyze how user characteristics affect filter bubbles. For unique tastes of users, we match the corresponding keywords of each genre to form \textit{Genre Preferences}. Then, for each genre preference, we calculate the average value of the $\mathsf{BEP}$ of all users with it. Figure~\ref{fig:bep_group_analysis} shows $\mathsf{BEP}$ of users with different \textit{Genre Preferences}. The results reveal the following observations:

\begin{itemize}
    \item Specifically, active users (higher activity level) demonstrate low $\mathsf{BEP}$ values, indicating that systems tend to restrict exposure even when users frequently interact with the platform. Similarly, users with narrow interests (lower diversity level) also experience more difficulty escaping the filter bubble, as indicated by lower $\mathsf{BEP}$ scores. These results suggest that user effort alone is not sufficient to overcome algorithmic confinement.

    \item Regarding the conformity level, there is no obvious correlation with $\mathsf{BEP}$. Considering the definition of conformity, this is in line with common sense.

    \item Furthermore, we observe notable differences in $\mathsf{BEP}$ across genre preferences. As shown in Figure~\ref{fig:bep_group_analysis} (d), users favoring niche or high-engagement genres such as thriller (Thr.), action (Act.), and crime (Cri.) exhibit significantly higher $\mathsf{BEP}$ than those favoring documentaries or animations, resembling a long-tail trend \cite{wang2023unified, yang2024hierarchical, li2025focalsam, wang2025localization}. This implies that the underlying content ecosystem also plays a role in how filter bubbles form and persist, reinforcing the importance of modeling both user behavior and item characteristics in filter bubble analysis.
\end{itemize}

\section{Conclusion}

This paper offers a novel perspective on understanding and mitigating filter bubbles in recommendation systems by introducing the metric of bubble escape potential ($\mathsf{BEP}$). Unlike traditional metrics that are entangled with user preference modeling, our metric provides a behavior-independent, quantitative approach to assess the severity of filter bubble. Through empirical analysis, we demonstrate how different recommendation systems vary in their tendency to create filter bubbles and explore the potential of random recommendation strategies to alleviate this issue. These findings advance understanding of the accuracy–bubble dilemma and provide a foundation for developing more inclusive, socially responsible recommendation systems.

\section{Acknowledgements}

This work was supported in part by National Natural Science Foundation of China: 62525212, 62236008, 62441232, U21B2038, U23B2051, 92370102, and 62502500, in part by Youth Innovation Promotion Association CAS, in part by the Strategic Priority Research Program of the Chinese Academy of Sciences, Grant No. XDB0680201, in part by the China National Postdoctoral Program for Innovative Talents under Grant BX20240384, in part by Beijing Natural Science Foundation under Grant No. L252144, in part by General Program of the Chinese Postdoctoral Science Foundation under Grant No. 2025M771558, and in part by the Fundamental Research Funds for the Central Universities.

\bibliography{aaai2026}

\end{document}